\documentclass[fleqn,twoside]{article}
\usepackage{espcrc2}

\usepackage{graphicx}

\newcommand{\AmS}{{\protect\the\textfont2
  A\kern-.1667em\lower.5ex\hbox{M}\kern-.125emS}}
\newcommand{\epe}{\epsilon^\prime/\epsilon}
\newcommand{\be}{\begin{equation}}
\newcommand{\ee}{\end{equation}}
\newcommand{\bea}{\begin{eqnarray}}
\newcommand{\eea}{\end{eqnarray}}

\newcommand{\etal}{{\it et al}.\ }

\title{Quenching effects in strong penguin contributions to $\epsilon'/\epsilon$}

\author{Jack Laiho$^{a}$\\ \bigskip $^{a}$Department of Physics, Princeton University,
Princeton, NJ\ \ 08544}

\begin{document}

\begin{abstract}
Quenching effects in strong penguin matrix elements are
investigated.  A lattice determination of $\alpha_q^{NS}$, the
constant that appears in the quenched ChPT relevant for the
lattice analysis of $K\to\pi\pi$ matrix elements, shows that this
constant is large.  The original RBC analysis of $Q_6$ matrix
elements is revisited in light of this result.  Also, the
numerical effects of choosing the singlet Golterman-Pallante
method of quenching $Q_6$ is investigated.
\end{abstract}
\vskip -.05in

\maketitle \vskip -0.5in
\section{Introduction}

There have been several recent lattice attempts to calculate
$\textrm{Re}(\epe)$, the direct \emph{CP} violating parameter in
$K \to \pi\pi$ decays.  These include the attempts using domain
wall fermions by the CP-PACS \cite{noaki} and RBC \cite{blum}
collaborations.  A notable feature of both of these calculations
is that their central values differ drastically from experiment,
though the approximations made were rather severe.  In particular,
the quenched approximation and leading order chiral perturbation
theory were employed in both calculations.

Although the quenched approximation is uncontrolled, where
quenched lattice results have been compared to experiment, at
least for simple quantities such as masses of flavored mesons and
decay constants, the agreement is at or better than $\sim 25\%$.
However, there is no good reason to think that this agreement
should hold for all low-energy hadronic phenomena. In fact there
is a particular difficulty with quenched $\epe$ that is not
present in other quenched calculations.

Since the original RBC $\epe$ result was reported, it was shown by
\cite{goltthree,goltfour} that there is an ambiguity in defining
the quenched strong penguin four-quark operators.  This ambiguity
affects the operator,

\bea\label{1} Q_{6}=\overline{s}_{a} \gamma_{\mu} (1-\gamma^{5})
d_{b} \sum_{q}
    \overline{q}_{b}\gamma^{\mu} (1+\gamma^{5}) q_{a}.
    \eea

\noindent whose $K\to\pi\pi$ matrix element has a large
contribution to $\epe$.  There are at least two ways of defining
this operator in the quenched theory, and we dub the calculations
done with these different definitions the quenched singlet and
non-singlet methods.  The non-singlet method corresponds to the
choice of the operator made in the standard (original RBC)
approach, while the singlet choice corresponds to the
Golterman-Pallante proposal.  It is not known a priori whether
either of these two methods gives results close to those of the
full theory, and debates on this matter are likely to be resolved
only when we have the results of dynamical calculations with the
physical number of light flavors. It is possible that an
appropriate linear combination of the two methods may give a
reasonable approximation to the full theory; see the contribution
of Norman Christ to these proceedings for such a proposal.

For the non-singlet method used (implicitly) in RBC's previous
work, it was shown that $K\to\pi\pi$ may have a large contribution
from an additional low energy constant (LEC), as suggested by the
large $N_{c}$ approximation, where $N_c$ is the number of colors
\cite{goltfive}.  Since for all $\Delta I=1/2$ amplitudes, the
leading order ChPT analysis employed in \cite{blum} was that of
the full theory, it seems desirable to revisit the analysis of
$Q_6$ matrix elements in light of the new results in quenched
ChPT.  We emphasize that this ambiguity is present only for the
strong penguin operators, and so does not affect the $Q_8$
contribution to $\epe$.  This quenching ambiguity also does not
affect Re $A_0$, Re $A_2$ and, therefore, the $\Delta I=1/2$ rule.

\section{Singlet and non-singlet methods}

We discuss the situation for $Q_6$ in order to illustrate the
subtlety involved in quenching the strong penguin operators. From
Eq (1) it can be seen that the right part of $Q_6$ is a sum over
active light flavors, so that in the full theory the right part is
a flavor singlet under the symmetry group $\textrm{SU}(3)_R$. In
the quenched theory one has at least two options. In the first
option one chooses to sum over all the quarks, including valence
and ghost.  In this case the right component of the operator
transforms as a singlet under the extended symmetry group; this is
called the quenched singlet option.

In the second option, one may choose to sum in Eq (\ref{1}) over
only the valence quarks.  In this case the operator can be
decomposed into two operators, one of which transforms as a
singlet under the irrep of the extended symmetry group, while the
other does not (rather, for $Q_6$, it transforms in the adjoint
representation); we choose to call this the quenched non-singlet
method.  In this method a new constant, $\alpha_q^{NS}$, appears
at leading order (LO), and a large $N_C$ estimate \cite{goltfive}
has suggested that this constant may be large.

\section{Measuring $\alpha_q^{NS}$}

\begin{figure}
\begin{center}
\includegraphics[scale=.4]{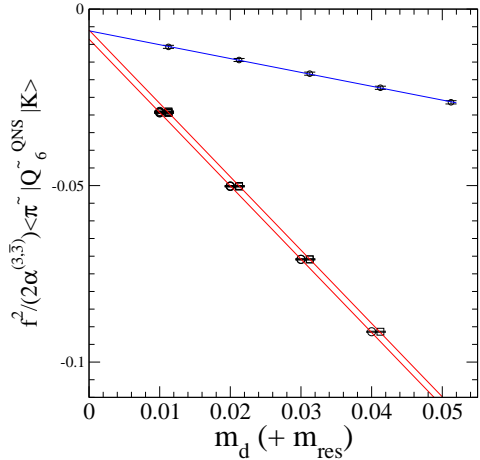}
\end{center}
\vskip -0.5in \caption{$(\langle
0|\tilde{Q}^{QNS}_6|\tilde{K}\rangle + \langle
0|Q^{QNS}_6|K\rangle)/\langle 0|\overline{s}\gamma_5 d|K\rangle$
as a function of $m_d$ (big circles) and $m_d+m_{res}$ (squares),
and $f^2/(2\alpha^{(3,\overline{3})})\langle
\tilde{\pi}|\tilde{Q}^{QNS}_6|K\rangle$ (small circles) as a
function of $m_d$ equal to the degenerate quark mass $m_f$.  The
pre-factor of $\langle \tilde{\pi}|\tilde{Q}^{QNS}_6|K\rangle$ was
chosen so that the two methods would have the same value in the
chiral limit according to ChPT. \label{fig8}} \vskip -0.3in
\end{figure}

It is difficult to numerically disentangle the $\alpha_q^{NS}$ log
term from the linear term and possible higher order effects where
it appears in $K\to 0$ \cite{goltfour}. It was for this reason
that \cite{goltfour} provided a matrix element to which
$\alpha_q^{NS}$ contributes at $O(p^0)$, so that it may be
obtained more readily in the chiral limit. This can be done if one
considers a matrix element where the ghost quarks can appear on
external lines.

Golterman and Pallante proposed looking at the matrix element,
$\widetilde{K}\to 0$, where $\alpha_q^{NS}$ can be obtained in the
chiral limit.  The $\widetilde{K}$ is a meson made of a quark
anti-ghost.  We have shown that there is a graded CPS symmetry
that is relevant for matrix elements with ghosts in external
states that is analogous to the usual CPS symmetry used to
constrain the power divergences of kaon decay amplitudes. Using
this symmetry, we have shown that the $\widetilde{K}\to 0$ matrix
element has a quadratic divergence. Also, because the chiral
symmetry of domain wall fermions with finite fifth dimension is
not exact, there is a contamination in the chiral limit of
$O(m_{res}/a^2)$ where $m_{res}$ is the contribution to the
fermion mass from chiral symmetry breaking.  Figure 1 shows that
the size of this ambiguity is of the same order of the chiral
limit we are trying to extract.

For this reason we have proposed an alternative method for
obtaining $\alpha_q^{N}$ that does not suffer from this problem.
Using the graded CPS symmetry we have shown that for degenerate
quark masses, the amplitude $K\to\widetilde{\pi}$ does not have
divergences.  This matrix element calculated on the same
configurations is plotted in Figure 1, and the value of
$\alpha_q^{NS}$ so obtained is in rough agreement with the large
$N_C$ calculation \cite{goltfive}, implying that this constant
must be taken into account in the RBC analysis.

\section{Effects of $\alpha_q^{NS}$ in the non-singlet analysis}

The LO ChPT prediction for the $K\to 0$ matrix element including
the $\alpha_q^{NS}$ does not produce a good fit to the data
(uncorrelated $\chi^2/\textrm{dof}\approx 30$), and it is possible
that higher order effects need to be included in the analysis. The
data appears quite linear, and cannot accommodate the large log in
the analytical expression. Unfortunately, the NLO contributions
for the non-singlet method are not currently known, so it is not
possible to do a systematic analysis.  The corrections to $K\to0$
for the singlet part of the $Q_6$ operator required a one-loop
calculation to go to NLO [$O(p^4)$] \cite{golt}. The non-singlet
contribution to $O(p^4)$, however, would require a two-loop
calculation, but this is not likely to be available anytime soon.

It is clearly difficult to separately resolve nonlinearities using
the numerical data if there is, in fact, a cancellation between
logarithmic and quadratic terms over the range of quark masses
used in our simulations.  Although there is reasonable agreement
between the results for the subtracted $K\to\pi$ matrix element
leaving out the heaviest mass data points in $\langle 0| Q_6 | K^0
\rangle$ versus including a quadratic term in the fit and using
all ten data points, we do not believe we have enough control over
the subtraction in order to obtain the needed LEC's from
$K\to\pi$.  We, therefore, cannot claim a result for $K\to\pi\pi$
in the non-singlet method.


\section{Results for the singlet method}

Although the quenched uncertainties make it unclear as to which
method is best to approximate the full theory, the singlet method
has the practical advantage that the NLO formulas are known,
making it a useful test of the whole formalism, including a
comparison to (quenched) continuum methods.  The quenched NLO
formulas do indeed produce good fits to the data. We are able to
obtain a subtracted $K\to\pi$ amplitude which is shown in Fig 2,
along with the original RBC data reported for subtracted $K\to\pi$
(where the $\alpha_q^{NS}$ term was not taken into account and a
linear fit was used), for comparison. The slope of this plot as a
function of $m_f$ gives the LEC needed for LO $K\to\pi\pi$
determinations, and one can see that the slope is enhanced by
roughly a factor of two in the singlet method over the original
RBC result.  On the other hand, a complete NLO fit shows that the
linear part of the singlet fit (the slope in the chiral limit) is
roughly parallel with the slope quoted for the original RBC data.
Thus, it is extremely important to have the NLO formulas as the
result can be affected by a factor of two or more.

\begin{figure}
\begin{center}
\includegraphics[scale=.4]{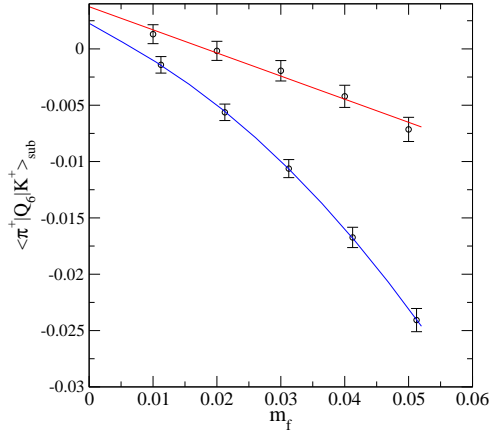}
\end{center}
\vskip -0.48in \caption{The lower data set is the matrix element
$\langle \pi^+| Q_6^{QS} |K^+ \rangle_{sub}$ where the divergent
piece has been subtracted. This was obtained using the singlet
method, and the fit is to the known NLO form.  The upper data set
is the original RBC result for $\langle \pi^+| Q_6 |K^+
\rangle_{sub}$ with a linear fit shown for comparison.
\label{fig14}} \vskip -0.34in
\end{figure}

\section{Outlook}

The large quenching effects show that dynamical calculations will
be necessary in order to significantly reduce this large
systematic uncertainty in $\epe$.  Also, if the reduction method
of \cite{bern} is used, it appears necessary to have NLO ChPT in
order to extract the needed LEC's from lattice data.

\end{document}